\documentclass[vecphys]{svmult}
\usepackage{paralist}
\usepackage{graphicx}
\usepackage{makeidx}
\usepackage{multicol}
\usepackage[bottom]{footmisc}
\usepackage{amssymb}
\usepackage{wasysym}

\makeindex

\begin{document}
\title*{Terrestrial Planet Formation in Binary Star Systems} 

\author{Elisa V. Quintana\inst{1} and Jack J. Lissauer\inst{2}}

\institute{SETI Institute, 515 N. Whisman Road, Mountain View, CA 94043, USA
\texttt{equintana@mail.arc.nasa.gov}
\and 
Space Science and Astrobiology Division 245-3, NASA Ames
  Research Center, Moffett Field, CA 94035, USA}


\titlerunning{Terrestrial Planet Formation in Binary Star Systems}
\authorrunning{Quintana et al.}




\maketitle

\section{Introduction}
\label{sec:intro}

More than half of all main sequence stars, and an even larger fraction
of pre-main sequence stars, reside in binary/multiple star systems
\cite{duq91, mat00}.  Numerical simulations of the collapse of
molecular cloud cores to form binary stars suggest that disks 
form within binary star systems \cite{bod00}.  The presence of disk
material has been indirectly observed around one or both components of
some young binary star systems \cite{mat00}.  Terrestrial planets and the cores of
giant planets are thought to form by an accretion process within a
disk of dust and gas \cite{saf69, lis93}, and therefore may be common
in binary star systems.  A lower limit of 30 (Jupiter mass) extrasolar planets have been detected in
so-called S-type orbits which encircle one member of a binary star
system.  Most of these extrasolar planets orbit stars whose stellar
companion is quite far away, but 3 are in systems with stellar
semimajor axes, $a_B$, of only $\sim$ 20 AU.  The effect of
the stellar companion on the formation of these planets remains
uncertain.  Terrestrial planets have yet to be detected in
P-type orbits (which encircle both components) of a main-sequence binary
star system, but close binaries are not included in precise Doppler
radial velocity search programs because of their complex and varying
spectra.  The presence of Earth-like exoplanets in main sequence single or binary star systems has yet to be determined, though projects that
focus on searching for extrasolar terrestrial planets are currently in
development\footnote{see www.kepler.arc.nasa.gov} and could begin to provide
insight within the next few years.

We have numerically simulated the final stages of terrestrial planet
formation in both S-type and P-type orbits within main-sequence binary
star systems using two symplectic integration algorithms that we
developed for this purpose \cite{cha02}.  Runs are performed using various values for the stellar masses
and orbital parameters in order to determine whether/where terrestrial
planets can form.  Multiple simulations are performed of each binary
star system under study to account for the chaotic nature of these
$N$-body systems, and we statistically compare the resulting planetary
systems formed to those that form in simulations of the
Sun-Jupiter-Saturn system which begin with near identical initial disk
conditions.  We show that (at least the final stages of) terrestrial planet formation can indeed take
place in a wide variety of binary star systems, and we have begun to
delineate the range of binary star parameter space (masses and orbits)
for which Earth-like planets may grow.  In this chapter, we present a
summary of the results of our simulations of planetary growth in
S-type orbits around each star in the $\alpha$ Centauri system (\S 2), 
in other `wide' binary star systems (\S 3), and in P-type orbits around both stars in close
short-period binary star systems (\S 4).  Details of the algorithms that we
helped to develop and the corresponding performance tests are presented in Chambers (\cite{cha02}) and in the previous chapter. Details of our research in terrestrial planet formation in binary
star systems are provided in \cite{qui02, qui03, qui04, lis04, qui06, qui07}.   The results from all of our simulations presented in this chapter (as well as simulations of planetary growth around single stars) can be scaled for different star and disk
parameters with the formulae presented in Appendix C of \cite{qui06}.  



\section{Model and Initial Conditions}

Our simulations are based on the conventional model of
planet formation in which terrestrial planets form via pair-wise
accretion of rocky bodies from within a disk of gas and dust that has
remained around a newly formed star \cite{saf69, lis93}.  For all of
our simulations, we assume that rocky Moon-to-Mars-sized
planetesimals/embryos have already accreted from within a disk of gas
and dust.  The initial conditions and
physical assumptions of the bodies in the `bimodal' disk are based upon
simulations of the final stages of terrestrial planet growth around
single stars \cite{cha01}.  Although this formulation is not complete, nor
definitive, it provides a model that reproduces our terrestrial planet
system (albeit with somewhat larger eccentricities) and can thus be
used as a reference point.  A comparison between our results and
those of the single-star accretion simulations will help to delineate the
effects of the binary star system on the accretion process despite our
approximations, e.g., the lack of very small bodies and the assumption
of perfectly inelastic collisions.

This initial bimodal disk mass distribution adopted from Chambers
\cite{cha01} is used for all of the simulations presented in this
chapter.  In this disk model, half of the disk mass is composed of 14
rocky embryos (each with a mass of 0.0933 times the mass of the Earth,
M$_{\oplus}$), while the remaining mass is distributed equally among
140 planetesimals (each with a mass of 0.00933 M$_{\oplus}$).  The
total disk mass is $\sim$ 2.6 M$_{\oplus}$. The profile of the surface
density has the form $a^{-3/2}$ (where $a$ is the semimajor axis),
normalized to 8 g/cm$^2$ at 1 AU, and follows from models of the
minimum mass solar nebula.  The bodies are distributed between 0.36 AU
and 2.05 AU, and the radius of each body is calculated assuming a
material density of 3 g cm$^{-3}$.  The embryos/planetesimals begin
with initial eccentricities $e \leq$ 0.01, inclinations $i \leq$
0.5$^{\circ}$, and specific initial orbital elements were chosen at
random from specified ranges; the same set of randomly selected values
was used for all simulations.

The evolution of the accreting bodies subject to gravitational
perturbations from both stars and to gravitational interactions and
completely inelastic collisions among the bodies is followed for 200
Myr -- 1 Gyr.  Because these $N$-body systems are chaotic in nature,
we performed multiple integrations of each system with a slight change
in the initial conditions of one body in the disk.  This tactic allows
us to sample the range of possible outcomes for effectively equivalent
initial conditions, and the result is a distribution of final
planetary systems.


\section{Planet Formation in the $\alpha$ Centauri AB Binary Star System}

We first examined the late stage of planet formation in the $\alpha$
Centauri system, which is comprised of a central binary consisting of
the G2 star $\alpha$ Cen A (1.1 M$_{\odot}$) and the K1 star $\alpha$
Cen B (0.91 M$_{\odot}$) \cite{qui02}.  The stars have an orbital
semimajor axis of 23.4 AU and an eccentricity of 0.52.  The M5 star
$\alpha$ Cen C (Proxima Centauri) is thought to orbit this pair, but
at a very large distance (12,000 AU), and is neglected in our
simulations.  Observations at the Anglo-Australian telescope imply
that no planet orbiting either star induces periodic velocity
variations as large as 2 m/s (G. Marcy, personal communication, 2006).
This upper bound, combined with dynamical stability calculations
\cite{wie97}, implies that any planet in an S-type orbit around either
component of the $\alpha$ Cen AB binary must have a mass less than
that of Saturn or orbit in a plane that is substantially inclined to
the line of sight to the system.  Herein we present the results of
simulations of the late stages of terrestrial planet growth around
$\alpha$ Cen A and $\alpha$ Cen B for various initial inclinations of
the circumstellar disk relative to the binary orbital plane.

In the majority of our simulations, the circumstellar disk is centered
around $\alpha$ Centauri A with $\alpha$ Cen B perturbing the system.
The initial inclination of the midplane of the disk, $i$, begins at
either 0$^{\circ}$, 15$^{\circ}$, 30$^{\circ}$, 45$^{\circ}$,
60$^{\circ}$, or 180$^{\circ}$ relative to the plane containing
$\alpha$ Cen A and B.  Although a stellar companion present during the
earlier stages of planet formation would likely force the planetesimal
disk into the plane of the binary orbit, many binary stars may
originate as unstable triple star systems which could produce a binary
star system with an accretion disk at a high relative inclination.  It
is also possible that a companion may have been captured around a
single star that posesses an accretion disk. The longitude of
periastron of the stellar companion is set to either 90$^{\circ}$ or
180$^{\circ}$ for each run.  We also performed a set of integrations
with the disk centered around $\alpha$ Cen B, with $\alpha$ Cen A
orbiting the system in the same plane and direction ($i$ =
0$^{\circ}$).  For comparison purposes, we performed a set of runs
which follow the evolution of the bimodal accreting disk around the
Sun with neither giant planets nor a stellar companion perturbing the
system.

Figure 1 shows the results from a simulation in which the disk is
centered around $\alpha$ Cen A and coplanar to the binary orbital
plane.  Each panel shows the eccentricity of each body in the disk as
a function of semimajor axis at the specified time, and the radius of
each symbol is proportional to the radius of the body that it
represents.  Within 100 Myr of the integration, five terrestrial
planets that are at least as massive as the planet Mercury have formed
around $\alpha$ Cen A, with a single planetesimals remaining in a
highly eccentric orbit.  The planetesimal is ejected from the system
soon thereafter (110 Myr), and the five terrestrial planets remain on
stable orbits within 2 AU for the remainder of the 200 Myr simulation.

The simulation shown in Figure 2 begins with nearly identical initial
conditions as the system in Figure 1, with the exception of a small (1
meter) shift in the initial position of one planetesimal near 1 AU.
Although the early evolution of the disk is qualitatively similar
among the two systems, they diverge with a Lyapunov time of order
$10^2$ years (\cite{qui07}).  In the simulation shown in Figure 2,
this divergence ultimately leads to the formation of an substantially
different planetary system: four terrestrial planets form within 1.8
AU of $\alpha$ Cen A.  Although these $N$-body simulations are highly
stochastic, there are clear trends in the final planetary systems
(number, masses, orbits etc.) that form in simulations with the same
binary star parameters (as will be shown further in \S 4 and 5).

Figure 3 shows the results of an integration in which the disk is
initially inclined by 15$^{\circ}$ relative to the binary orbital
plane.  With a higher initial disk inclination (than the simulations
shown in the previous two figures), the bodies in the disk are more
dynamically excited, and more mass is lost from the system.  In this
case, nearly 15\% more mass is lost than the $i$ = 0$^{\circ}$
simulations, and three terrestrial planets (as well as a single planetesimal)
remain by the end of the integration.

We performed a total of 16 simulations of terrestrial planet growth
around $\alpha$ Cen A in which the midplane of the disk was initially
inclined by 30$^{\circ}$ or less relative to the binary orbital plane.
In these simulations, when the bodies in the disk began in prograde
orbits, from 3 -- 5 terrestrial planets formed around $\alpha$ Cen A.
Slightly more formed, from 4 -- 5, when $i$ = 180$^{\circ}$ relative
to the binary plane.  From 2 -- 4 planets formed in a disk centered
around $\alpha$ Cen B, with $\alpha$ Cen A perturbing the system in
the same plane.  The final planetary systems that form in all of these simulations are shown in Figures 8 -- 10 in \cite{qui02}.  The distribution of final terrestrial planet systems
in the aforementioned cases is quite similar to that produced by
calculations of terrestrial planet growth in the Sun-Jupiter-Saturn
system.

In contrast, terrestrial planet growth around a star lacking both
stellar and giant planet companions is slower and extends to larger
semimajor axis for the same initial disk of planetary embryos
(see Figure 12 in \cite{qui02}).  In systems with the accreting disk initially inclined
at 45$^{\circ}$ to the binary plane, from 2 -- 5 planets formed,
despite the fact that more than half of the disk mass was scattered
into the central star.  When the disk was inclined by 60$^{\circ}$,
the stability of the planetary embryos decreased dramatically, and
almost all of the planetary embryos and planetesimals were lost from
these systems (e.g., Figure 6 in \cite{qui02}).  Figures of the
temporal evolution of each of our $\alpha$ Cen simulations can be
found in \cite{qui02, qui04}.

\section{S-type Orbits in Other `Wide' Binary Star Systems} 

In this section we present the results from a survey ($\sim$ 120
numerical simulations) on the effects of a stellar companion on the
final stages of terrestrial planet formation in S-type orbits around
one component of a binary star system \cite{qui07}.  We examine binary
star systems with stellar mass ratios $\mu \equiv M_C / (M_{\star} +
M_C)$ = 1/3, 1/2, or 2/3, where $M_{\star}$ is the mass of the star
around which the protoplanetary disk is situated, and $M_C$ is the
mass of the companion.  The majority of our simulations begin with
equal mass stars ($\mu$ = 1/2) of either $M_{\star}$ = $M_C$ = 0.5
M$_{\odot}$ (Set A) or $M_{\star}$ = $M_C$ = 1 M$_{\odot}$ (Set B).
Simulations were also performed with a more massive `primary' star
(the one the disk is centered around), $M_{\star}$ = 1 M$_{\odot}$ and
$M_C$ = 0.5 M$_{\odot}$ ($\mu$ = 1/3, Set C), and also with a smaller
`primary' star of $M_{\star}$ = 0.5 M$_{\odot}$ and a more massive
companion $M_C$ = 1 M$_{\odot}$ ($\mu$ = 2/3, Set D).  The stellar
semimajor axis, $a_B$, and binary eccentricity, $e_B$, are chosen such
that the binary periastron takes one of the three values, $q_B$ = 5
AU, 7.5 AU, or 10 AU.  Note that binary systems with much wider
periastra would have little effect on terrestrial planet formation,
whereas systems with significantly smaller periastra would completely
destroy the initial disk of planetesimals.  The binary stars are
separated by $a_B$ = 10 AU, 13$\frac{1}{3}$ AU, 20 AU, or 40 AU, and
the eccentricities are varied in the range 0 $\leq e_B \leq$ 0.875.
The largest semimajor axis for which particles can be stable in any of
the systems that we explore is 2.6 AU; we therefore omit giant planets
(those in the Solar System in orbit beyond 5 AU) from our
integrations.

For our accretion simulations, our exploration of parameter space has
two coupled goals.  On one hand, we want to determine the effects of
the binary orbital elements on the final terrestrial planet systems
produced. On the other hand, for a given binary configuration, we want
to explore the distribution of possible resulting planetary systems
(where the results must be described in terms of a distribution due to
the sensitive dependence on the initial conditions).  We have
performed from 3 -- 30 integrations for each wide binary star
configuration ($\mu$, $a_B$, and $e_B$) considered herein, with small
differences in the initial conditions: a single planetesimal is moved
forward along its orbit by a small amount (1 -- 9 meters) in an orbit
near either 0.5, 1, or 1.5 AU, prior to the integration.  Ideally, of
course, one would perform larger numbers of integrations to more fully
sample the distributions of results, but computer resources limit our
sample size.  Figures (similar to Figures 1 -- 3) of the evolution of
most wide binary star systems that we simulated can be found in
\cite{lis04, qui04, qui07}.

The stellar mass ratio, $\mu$, and the periastron distance $q_B$ strongly
influence where terrestrial planets can form in `wide' binary star systems.
The effect of $q_B$ on the distribution of final planetary system
parameters (i.e., number, masses, etc.) is demonstrated in Figures 
4 -- 6.  The semimajor axis of the outermost planet can be used as a
measure of the size of the terrestial planet system.  Figure 4 shows
the distribution of the semimajor axis of the outermost final planet
formed in each simulation for systems with $q_B$ = 5 AU (top panel),
7.5 AU, (middle panel), and 10 AU (lower panel).  Note that twice as
many integrations have been performed in Set B (shown in light gray) as in
Set A (dashed bars).  Figure 4 shows a clear trend: as the binary
periastron increases, the distribution of semimajor axes (of the
outermost planet) becomes wider and its expectation value shifts to
larger values.  The distributions of the total number of final planets
formed are shown in Figure 5 for simulations with $q_B$ = 5 AU, 7.5
AU, and 10 AU.  In general, a smaller binary periastron results in a
larger percentage of mass loss, and a smaller number of final planets.
From 1 -- 3 planets formed in all systems with $q_B$ = 5 AU, 1 -- 5
planets remained in systems with $q_B$ = 7.5 AU, and 1 -- 6 planets
formed in all systems with $q_B$ = 10 AU.  The range in the number of
possible planets grows with increasing binary periastron; similarly,
the average number of planets formed in the simulations is an
increasing function of $q_B$.  In the $q_B$ = 7.5 AU set with equal
mass stars of 1 M$_{\odot}$, an average of 2.8 planets
formed in the distribution of our largest set of 30 integrations.  The
distribution extends farther out if the perturbing star is smaller
relative to the central star for a given stellar mass ratio, and
slightly farther out when the stars are more massive relative to the
disk.  Figure 6 shows the distribution of final planetary masses (in
units of the Earth's mass, M$_{\oplus}$) formed in systems with $q_B$
= 5 AU, 7.5 AU, and 10 AU.  The median mass of the final planets
doesn't depend greatly on $q_B$.  This result suggests that planet
formation remains quite efficient in the stable regions, but that the
size of the stable region shrinks as $q_B$ gets smaller. This trend is
consistent with the decline in the number of planets seen in Figure 5.
When the periastron value becomes as small as 5 AU, planets only form
within 1 AU, and the mass distribution tilts toward $m_p <$ M$_{\oplus}$, i.e., the formation of Earth-like planets is compromised.

\section{P-type Orbits Within Close Binary Star Systems}
Herein we present results from simulations of the late stages of
terrestrial planet formation within a circumbinary disk surrounding
various short-period binary systems.  For a more extensive discussion,
see Quintana \& Lissauer (\cite{qui06}).  The combined mass of the
binary stars is equal to 1 M$_\odot$ in all of these simulations, with
the stellar mass ratio $\mu$ equal to either 0.2 or 0.5.  Binary star
separations in the range $a_B$ = 0.05 AU -- 0.4 AU are examined, while
$e_B$ begins at 0, 1/3, 0.5, or 0.8 such that the stellar apastron
$Q_B \equiv a_B(1 + e_B)$ is 0.05 AU $\leq Q_B \leq$ 0.4 AU.  For most
of the simulations, the midplane of the circumbinary disk begins
coplanar to the stellar orbit, but for one set of binary star
parameters a relative inclination of $i$ = 30$^{\circ}$ is
investigated.  The initial disk of planetary embryos/planetesimals is
the same as that used for simulating accretion within our Solar System
\cite{cha01}, in the $\alpha$ Centauri AB system (\S 3), and in wide
binary star systems (\S 4).  Giant planets with masses and initial
orbits equal to those of Jupiter and Saturn at the present epoch are
included in the simulations, as they are in most simulations of the
late stages of terrestrial planet accumulation in our Solar System.
We use a `close-binary' algorithm which follows the accretion
evolution of each body in the disk relative to the center of mass of
the binary star system.  To account for the stochastic nature of these
simulations, each binary star system under study is simulated five or
six times with slightly different initial conditions for the
circumbinary disk.  We statistically compared our results to a large
set ($>$ 30) of simulations of the Sun-Jupiter-Saturn system that
began with virtually the same initial disk mass distribution
\cite{cha01, qui06}.

Figure 7 shows the evolution of the circumbinary disk centered around
two 0.5 M$_{\odot}$ binary stars with $a_B$ = 0.1 AU and in a circular
orbit.  The initial disk is the same as that shown in the first panel
of Figure 1, planetary embryos and planetesimals are represented by
circles whose sizes are proportional to the physical sizes of the
bodies, and the locations of the circles show the orbital semimajor
axes and eccentricities of the represented bodies relative to center
of mass of the binary stars. The perturbations on the inner edge of
the disk are apparent within the first million years, and in this case
five terrestrial planets have formed within 100 Myr, and continue on
stable orbits for the remainder of the 500 Myr integration.  Figures 8
and 9 show the results from two simulations with equal (0.5
M$_{\odot}$) mass stars.  In Figure 8 the stellar separation is $a_B$
= 0.2 AU and $e_B$ = 0.5 ($Q_B$ = 0.3), while in Figure 9 the stars
are separated by $a_B$ = 0.3 AU and have an eccentricity of $e_B$ =
1/3 ($Q_B$ = 0.4).  In both simulations, more than 70\% of the initial
disk mass was lost from the system within the first 50 Myr, and each
simulation resulted in a single planet that is at least as massive as
the planet Mercury.  The evolution of most of the close binary star
simulations are presented in \cite{lis04, qui04, qui06}, and
\cite{qui06} provides a statistical comparison of of the planets that
form around each binary star configuration to those that form around
the Sun (with giant planets included in each case).

Figure 10 show the distributions in the semimajor axis of the
innermost planets that formed for systems that began on initially
circular orbits (gray bars) and for those in which the binary star
eccentricity ranged from 1/3 $\geq e_B \geq$ 0.8.  For the
zero-eccentricity simulations, there is a clear separation at $\sim$ 1
AU among systems with $Q_B \leq$ 0.2 AU and $Q_B \geq$ 0.3 AU.  The
distributions of the innermost semimajor axes of the planets formed in
binary stars that began with higher eccentricities are wider, as each
set of these simulations resulted in at least one system with 1 -- 2
final planets, as shown in Figure 11.  Figure 12 shows the
distributions of the final masses of the planets formed among binary
star systems with different apastron values.  Earth-mass planets only
formed in our simulations in which $Q_B \leq$ 0.2 AU.

In summary, close binary stars with maximum separations $Q_B$ $\leq$
0.2 AU and small $e_B$ had little effect on the accreting bodies, and
in most of these simulations terrestrial planets formed over
essentially the entire range of the initial disk mass distribution
(and even beyond 2 AU in many cases).  The stellar perturbations cause
orbits to precess, thereby moving secular resonances out of the inner
asteroid belt, allowing terrestrial planets to form from our initially
compact disk and remain in stable orbits as far as 2.98 AU from the
center of mass of the binary stars.  The effects of the stellar
perturbations on the inner edge of the planetesimal disk became
evident in systems with larger $a_B$ (and $Q_B \gtrsim$ 0.3 AU) and in
most of the simulations with $e_B \textgreater$ 0.  Terrestrial-mass
planets can still form around binary stars with nonzero eccentricity,
but the planetary systems tend to be sparcer and more diverse.  Binary
stars with $Q_B \gtrsim$ 0.3 AU perturb the accreting disk such that the
formation of Earth-like planets near 1 AU is unlikely.  Despite these
constraints, at least one terrestrial planet (at least as massive as
the planet Mercury) formed in each of our simulations.

\section{Conclusions}
Our exploration of parameter space shows how binary orbital parameters
affect terrestrial planet formation.  We find that the presence of a
binary companion of order 10 AU away acts to limit the number of
terrestrial planets and the spatial extent of the terrestrial planet
region around one member of a binary star system, as shown by Figures
3 -- 5.  To leading order, the periastron value $q_B$ is the most
important parameter in determining binary effects on planetary
outcomes in wide-binary star systems (more predictive than $a_B$ or
$e_B$ alone), whereas $Q_B$ is the most influential parameter for
accretion within circumbinary disks.  In our ensemble of $>$ 100 wide
binary star simulations that began with equal mass stars, from 1 -- 6
planets formed with semimajor axes $\lesssim$ 2.2 AU of the central
star in binary systems with $q_B$ = 10 AU, from 1 -- 5 planets formed
within 1.7 AU for systems with $q_B$ = 7.5 AU, and from 1 -- 3 planets
formed within 0.9 AU when $q_B$ = 5 AU.  For a given
binary periastron $q_B$, fewer planets tend to form in binary systems
with larger values of ($a_B$, $e_B$), as shown in Figure 5.

Binary companions also limit the extent of the terrestrial planet
region in nascent planetary systems orbiting one member of the stellar pair. As shown in Figures 4 -- 6,
wider binaries allow for larger systems of terrestrial planets.  Although the binary periastron is the most important variable in determining
the extent of the final system of terrestrial planets (as measured by
the semimajor axis of the outermost planet), for a given
periastron, the sizes of the terrestrial planet systems show a wide
distribution.  In these simulations, the initial disk of planetesimals
extends out to 2 AU, so we do not expect terrestrial planets to form
much beyond this radius. For binary periastron $q_B$ = 10 AU, the
semimajor axis of the outermost planet typically lies near 2 AU, i.e.,
the system explores the entire available parameter space for planet
formation. Since these results were obtained with equal mass stars
(including those with $M$ = 1.0 M$_{\odot}$), we conclude that the
constraint $q_B \gtrsim 10$ AU is sufficient for binaries to leave
terrestrial planet systems unperturbed.  With smaller binary
periastron values, the resulting extent of the terrestrial planet
region is diminished. When binary periastron decreases to 5 AU, the
typical system extends only out to $a_p \sim$ 0.75 AU and no system
has a planet with semimajor axis beyond 0.9 AU (but note that we did
not perform simulations with $q_B =$ 5 AU and small $e_B$).

While the number of forming planets and their range of orbits is
restricted by binary companions, the masses and eccentricities of
those planets are much less affected.  The distribution of planet
masses is nearly independent of binary periastron (see Figure 6),
although the wider binaries allow for a few slightly more massive
terrestrial planets to form. Finally, we note that the time scales
required for terrestrial planet formation in these systems lie in the
range 50 -- 200 Myr, consistent with previous findings \cite{cha01,
  qui02, qui06}, and largely independent of the binary
properties. This result is not unexpected, as the clock for the
accumulation of planetesimals is set by their orbit time and masses
\cite{saf69}, and not by the binary orbital period.

This work has important implications regarding the question of what
fraction of stars might harbor terrestrial planetary systems.  The
majority of solar-type stars live in binary systems, and as shown in this chapter, binary
companions can disrupt both the formation of terrestrial planets and
their long term prospects for stability.  Approximately half of the
known binary systems are wide enough (in this context, having
sufficiently large values of periastron) so that Earth-like planets
can remain stable over the entire 4.6 Gyr age of our Solar System
\cite{dav03, fat06}. For the system to be stable
out to the distance of Mars's orbit, the binary periastron $q_B$ must
be greater than about 7 AU, and about half of the observed binaries
have $q_B > 7$ AU.  Our work on the formation of terrestrial planets
shows similar trends.  When the periastron of the binary is larger
than about $q_B$ = 10 AU, even for the case of equal mass stars,
terrestrial planets can form over essentially the entire range of
orbits allowed for single stars (out to the edge of the initial
planetessimal disk at 2 AU). When periastron $q_B < 10$ AU, however,
the distributions of planetary orbital parameters are strongly
affected by the presence of the binary companion (see Figures 7 --
12).  Specifically, the number of terrestrial planets and the spatial
extent of the terrestrial planet region both decrease with decreasing
binary periastron.  When the periastron value becomes as small as 5
AU, planets no longer form with $a = 1$ AU orbits and the mass
distribution tilts toward $m_p < 1 M_{\oplus}$, i.e., the formation of
Earth-like planets is compromised.  

Given the enormous range of orbital parameter space
sampled by known binary systems, from contact binaries to separations
of nearly a parsec, the range of periastron where terrestrial planet
formation is affected is quite similar to the range of periastron
where the stability of Earth-like planets is compromised. As a result,
$\sim$ 40 -- 50\% of binaries are wide enough to allow both the
formation and the long term stability of Earth-like planets in S-type
orbits encircling one of the stars.  Furthermore, approximately 10\% of
main sequence binaries are close enough to allow the formation and
long-term stability of terrestrial planets in P-type circumbinary
orbits (\cite{dav03, qui06}). Given that the
galaxy contains more than 100 billion star systems, and that roughly half
remain viable for the formation and maintenence of Earth-like planets,
a large number of systems remain habitable based on the dynamic
considerations of this research.

\section{Acknowledgements}
We thank Michael J. Way for providing additional CPUs at NASA ARC.
E.V.Q. received support in various stages of this research from NASA
GSRP, NAS/NRC and NASA NPP fellowships, and the University of Michigan
through the Michigan Center for Theoretical Physics (MCTP).  J.J.L. is
supported in part by the NASA Astrobiology Institute under the NASA
Ames Investigation ``Linking our Origins to our Future''.

%

{}

\clearpage

\begin{figure}[!t]\centering
  \includegraphics[width=1\columnwidth]{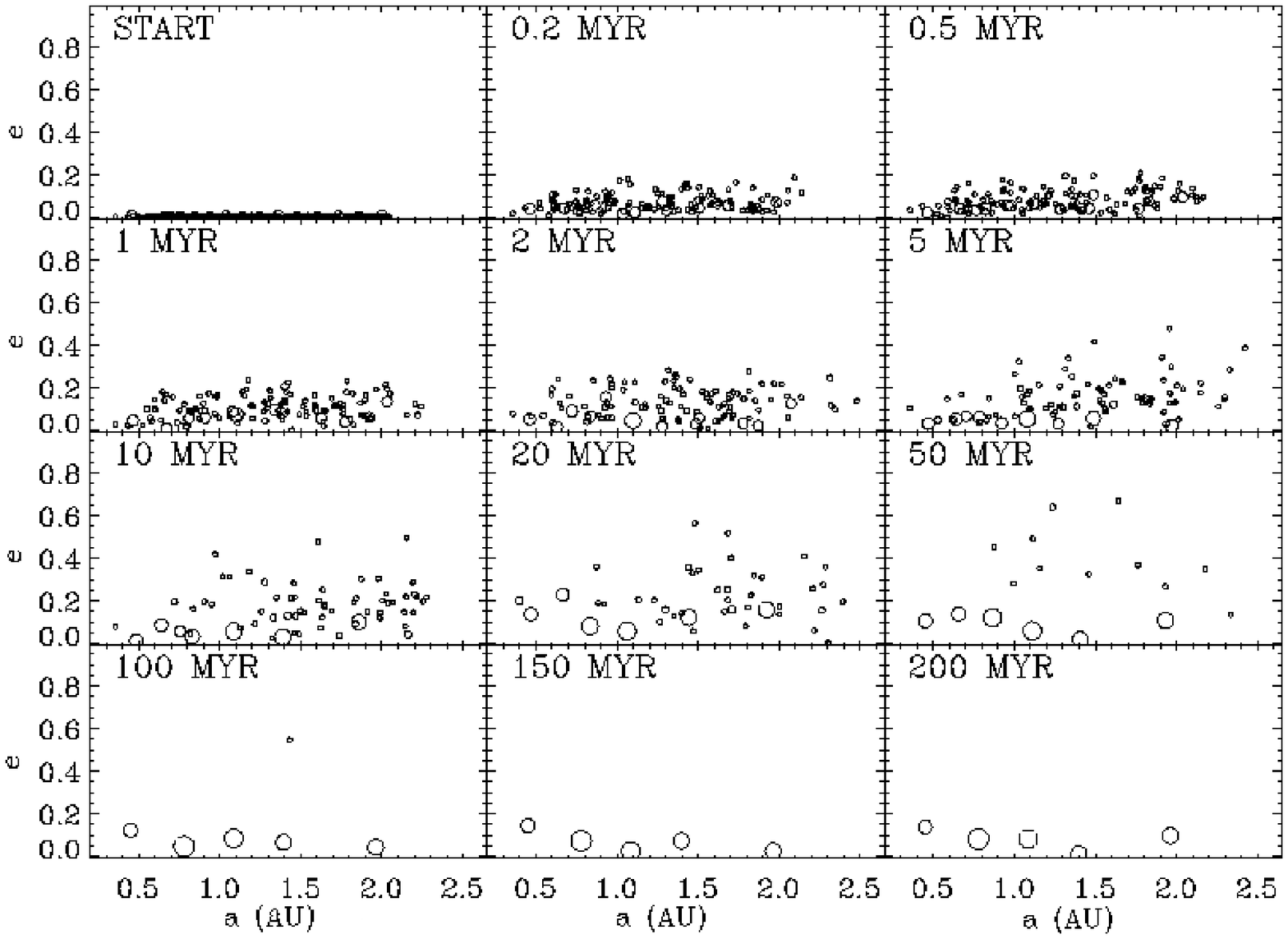}
  \caption{The temporal evolution of planetary embryos/planetesimals
    in a circumstellar disk centered around $\alpha$ Cen A and
    coplanar with the binary orbital plane (simulation AC$i0\_3$ in
    \cite{qui02}).  The radius of each symbol is proportional to the
    radius of the body that it represents, and the eccentricities are
    displayed as a function of semimajor axis.  By the end of the 200
    Myr integration, five terrestrial planets have formed within 2 AU
    of $\alpha$ Cen A, accumulating $\sim$ 89\% of the initial disk
    mass.  Figures 1 and 2 from \cite{qui02} show the results of two simulations that begin with the same stellar parameters (runs AC$i0\_1$ and AC$i0\_2$), but with slightly different initial disk conditions.}
\end{figure}

\begin{figure}[!t]\centering
  \includegraphics[width=1\columnwidth]{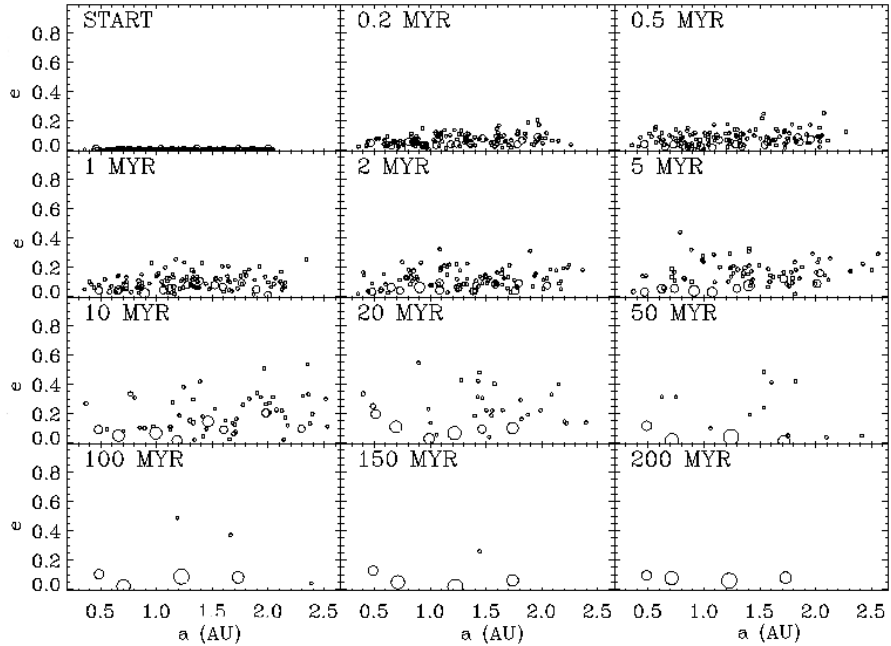}
  \caption{The temporal evolution of virtually the same circumstellar
    disk around $\alpha$ Cen A as that shown in Figure 1, but in this
    case a single planetesimal near 1 AU is shifted by a small amount
    (1 meter along its orbit) prior to the integration (simulation
    AC$i0\_4$ in \cite{qui02}).  The stellar parameters and all other
    disk initial conditions are identical to the system in Figure 1.
    The dynamics of the disk are generally the same in the earlier
    stages of the simulations shown here and in Figure 1, as the more
    massive embryos orbit with low eccentricities whereas the
    planetesimals are dynamically excited to much higher values.  The
    stochastic nature of these $N$-body systems is evident, however,
    in the final planetary system that formed, and demonstrates the
    sensitive dependence of the outcome on the initial conditions.
    Four terrestrial planets, comprised of $\sim$ 88\% of the initial
    disk mass, remain within 1.7 AU of $\alpha$ Cen A.  Figures 1 and 2 from \cite{qui02} show the results of two simulations that begin with the same stellar parameters (runs AC$i0\_1$ and AC$i0\_2$), but with slightly different initial disk conditions.}
\end{figure}

\begin{figure}[!t]\centering
  \includegraphics[width=1\columnwidth]{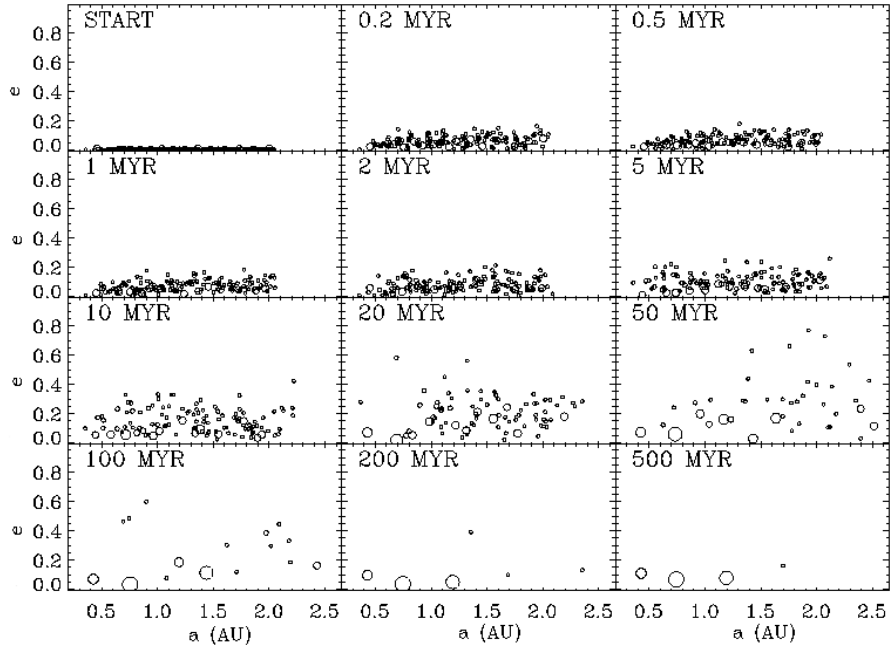}
  \caption{The temporal evolution of our standard circumstellar disk
    centered around $\alpha$ Cen A and initially inclined by
    15$^{\circ}$ to the binary orbital plane (simulation AC$i15\_2$ in
    \cite{qui02}).  The eccentricities of the embryos/planetesimals are
    displayed as a function of semimajor axis, and the radius of each
    symbol is proportional to the radius of each body that it
    represents.  In this case, three terrestrial planets formed within
    $\sim$ 1.2 AU, and a single planetesimal remained exterior to
    these planets at $\sim$ 1.7 AU, all composed of $\sim$ 74\% of the
    initial disk mass.  Similar evolution plots for all of the
    simulations involving $\alpha$ Cen A and B can be found in
    \cite{qui02, qui03, qui04, lis04}}
\end{figure}

\begin{figure}[!t]\centering
\includegraphics[angle=270, width=1\columnwidth]{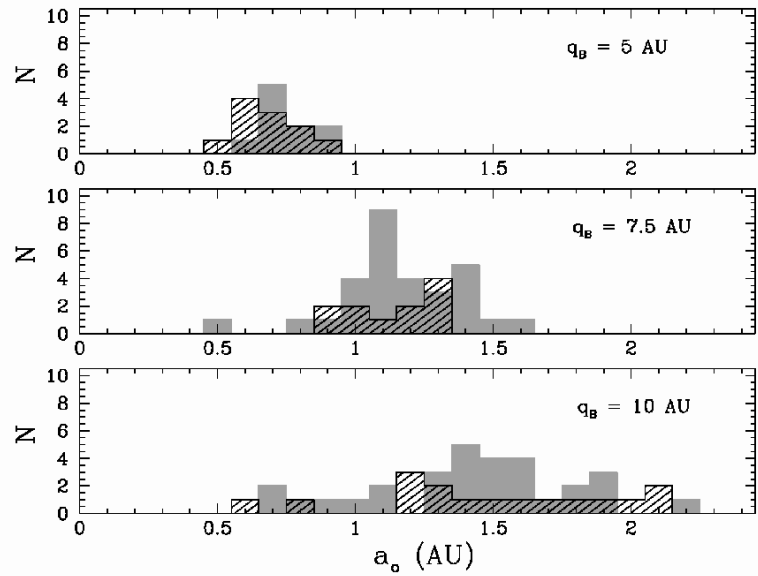} 
\caption{The distribution of the semimajor axis of the outermost final
  planet, $a_0$, formed For binary star systems with $q_B$ = 5 AU (top panel),
  $q_B$ = 7.5 AU (middle panel), and $q_B$ = 10 AU (bottom panel).
  The light gray bars represent simulations from Set A with
  $M_{\star}$ = $M_C$ = 0.5 M$_{\odot}$, whereas the dashed bars
  represent systems from Set B with $M_{\star}$ = $M_C$ = 1.0
  M$_{\odot}$.  Although the semimajor axes extend to larger values in
  binary systems with larger periastron, the inner edge of the
  distribution is roughly determined by the inner edge of the initial
  disk of embryos, as in the single star case, i.e., the presence of
  different stellar companions has a minimal effect on the inner
  terrestrial region.  Note that figures of these distributions (and
  also those shown in Figures 6 and 7) that include the results from
  simulations that began with unequal stars ($\mu \ne$ 0.5) are
  presented in \cite{qui07}.}
\end{figure}

\begin{figure}[!t]\centering
\includegraphics[angle=270, width=1\columnwidth]{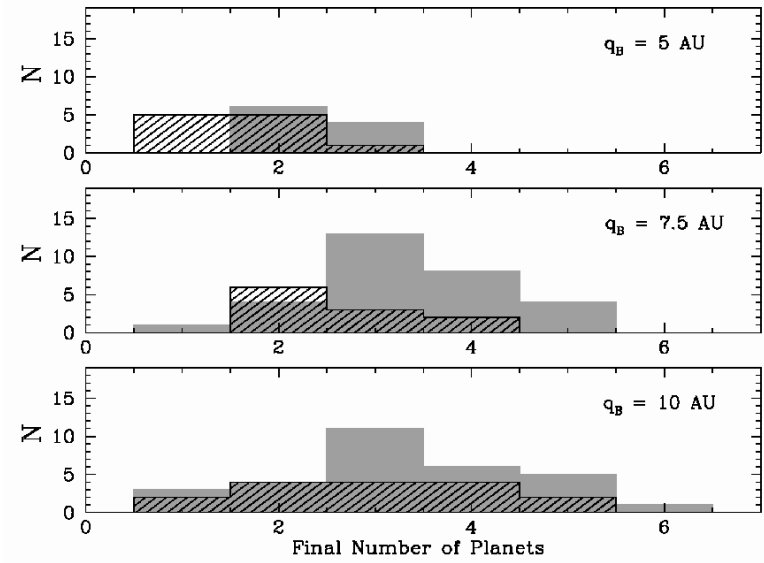}
\caption{Distributions of the number of final planets formed for
  binary star systems with $q_B$ = 5 AU (top panel), $q_B$ = 7.5 AU
  (middle panel), and $q_B$ = 10 AU (bottom panel).  The bar types correspond
  to the different sets of runs as described in Figure 4.  The
  typical number of final planets clearly increases in systems with
  larger stellar periastron, and also when the companion star is less
  massive than the primary (for a given stellar mass ratio).}
\end{figure}
  
\begin{figure}[!t]\centering 
\includegraphics[angle=270, width=1\columnwidth]{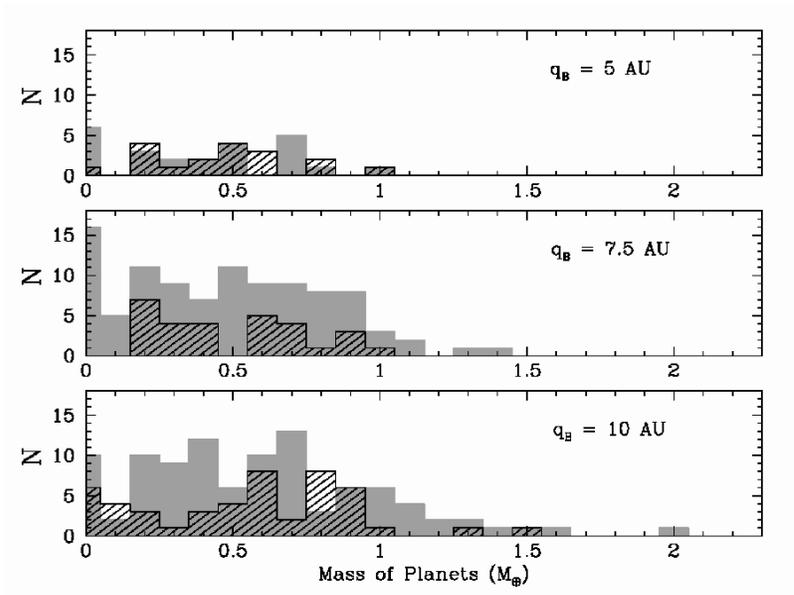}
\caption{Distributions of the final masses of planets formed within
  binary star systems with $q_B$ = 5 AU (top panel), $q_B$ = 7.5 AU
  (middle panel), and $q_B$ = 10 AU (bottom panel).  The bar types correspond
  to the different sets of runs as described in Figure 4.  Although the
  size of the stable region shrinks as $q_B$ gets smaller, the median
  mass of the final planets does not vary greatly for a given $q_B$,
  suggesting that planet formation remains efficient in the
  stable regions.}
\end{figure}

\begin{figure}[!t]\centering
\includegraphics[width=1\columnwidth]{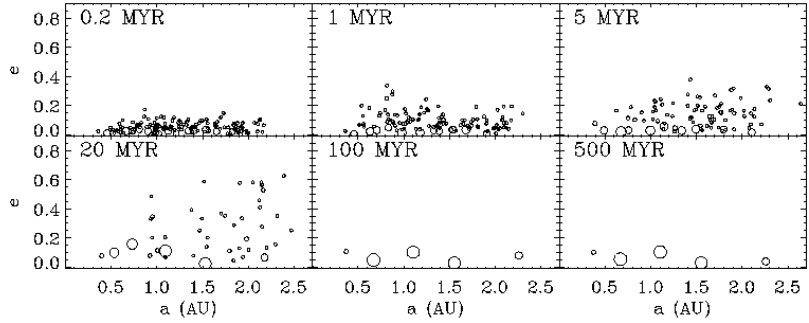}
\caption{The temporal evolution of the circumbinary disk around binary
  stars with $a_B$ = 0.1 AU, $e_B$ = 0, and equal mass stars of
  $M_{*}$ = 0.5 M$_{\odot}$.  Jupiter- and Saturn-like planets are
  also included.  The planetary embryos and planetesimals are
  represented by circles whose sizes are proportional to the physical
  sizes of the bodies.  The locations of the circles show the orbital
  semimajor axes and eccentricities of the represented bodies relative
  to center of mass of the binary stars.  The initially dynamically
  cold disk heats up during the first 10 Myr, especially in the outer
  region, where the perturbations of the single giant planet included
  in this simulation are the greatest.  By 100 Myr into the
  simulation, five planets on low eccentricity orbits have formed and
  survive for the remainder of the simulation.}
\end{figure}

\begin{figure}[!t]\centering
\includegraphics[width=1\columnwidth]{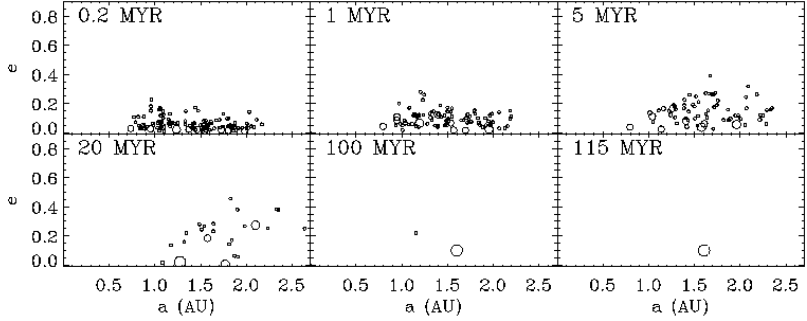}
\caption{
The temporal evolution of simulation that begins with a
  larger stellar separation ($a_B$ = 0.2 AU) and a higher binary
  eccentricity ($e_B$ = 0.5 AU) as the system shown in Figures 6.  
  The binary system is composed of
  equal mass stars of $M_{*}$ = 0.5 M$_{\odot}$, and Jupiter and Saturn are
  included.  The planetary embryos and planetesimals are
  represented by circles whose sizes are proportional to the physical
  sizes of the bodies as in Figure 6.  The locations of the circles
  show the orbital semimajor axes and eccentricities of the
  represented bodies relative to center of mass of the binary stars.
  The last ejection occurs at 115 Myr, in which only one planet
  remains in the system at $\sim$ 1.6 AU.}
\end{figure}

\begin{figure}[!t]\centering
\includegraphics[width=1\columnwidth]{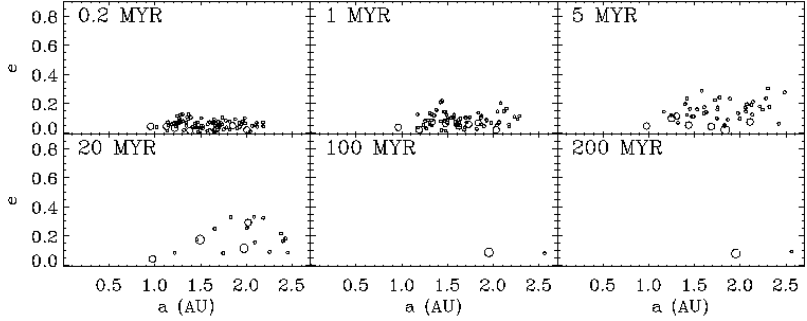}
\caption{This figure shows the early evolution of the circumbinary
  disk around binary stars with $a_B$ = 0.3 AU, $e_B$ = 1/3. The effect of the stellar companion is apparent in the first
  panel where the inner part of the disk is already substantially
  excited by 200,000 yr.  Eccentricities remain high throughout the
  evolution, and by 100 Myr only one planet more massive than the
  planet Mercury has formed in the terrestrial planet zone, and a
  single planetesimal remains beyond 2.5 AU.}
\end{figure}

\begin{figure}[!t]\centering
\includegraphics[angle=270, width=1\columnwidth]{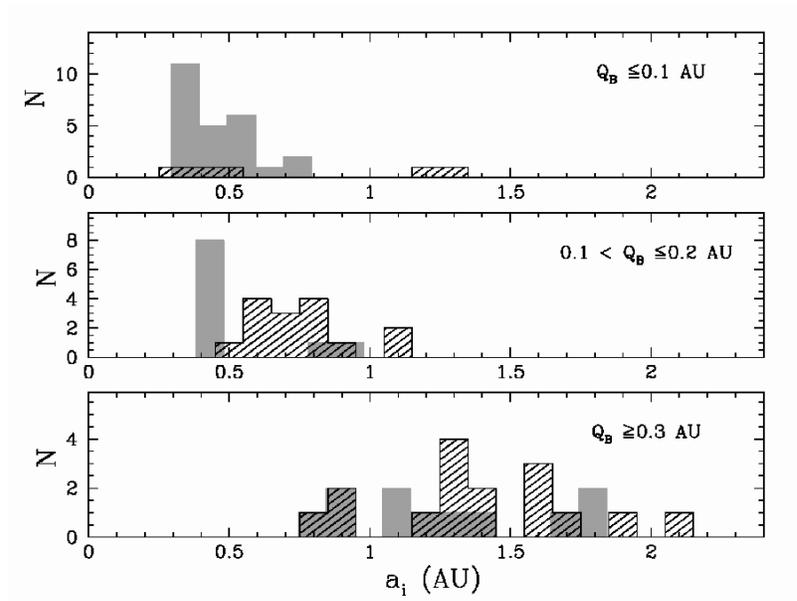} 
\caption{Distributions of the semimajor axis of the innermost final
  planet, $a_i$, formed in binary star systems with $Q_B \leq$ 0.1 AU
  (top panel), 0.1 $< Q_B \leq$ 0.2 AU (middle panel), and $Q_B \geq$
  0.3 AU (bottom panel).  The light gray bars represent simulations in
  which the binary stars began on circular ($i$ = 0$^{\circ}$) orbits,
  and the dashed bars represent systems with 1/3 $\geq e_B \geq$ 0.8.}
\end{figure}

\begin{figure}[!t]\centering
\includegraphics[angle=270, width=1\columnwidth]{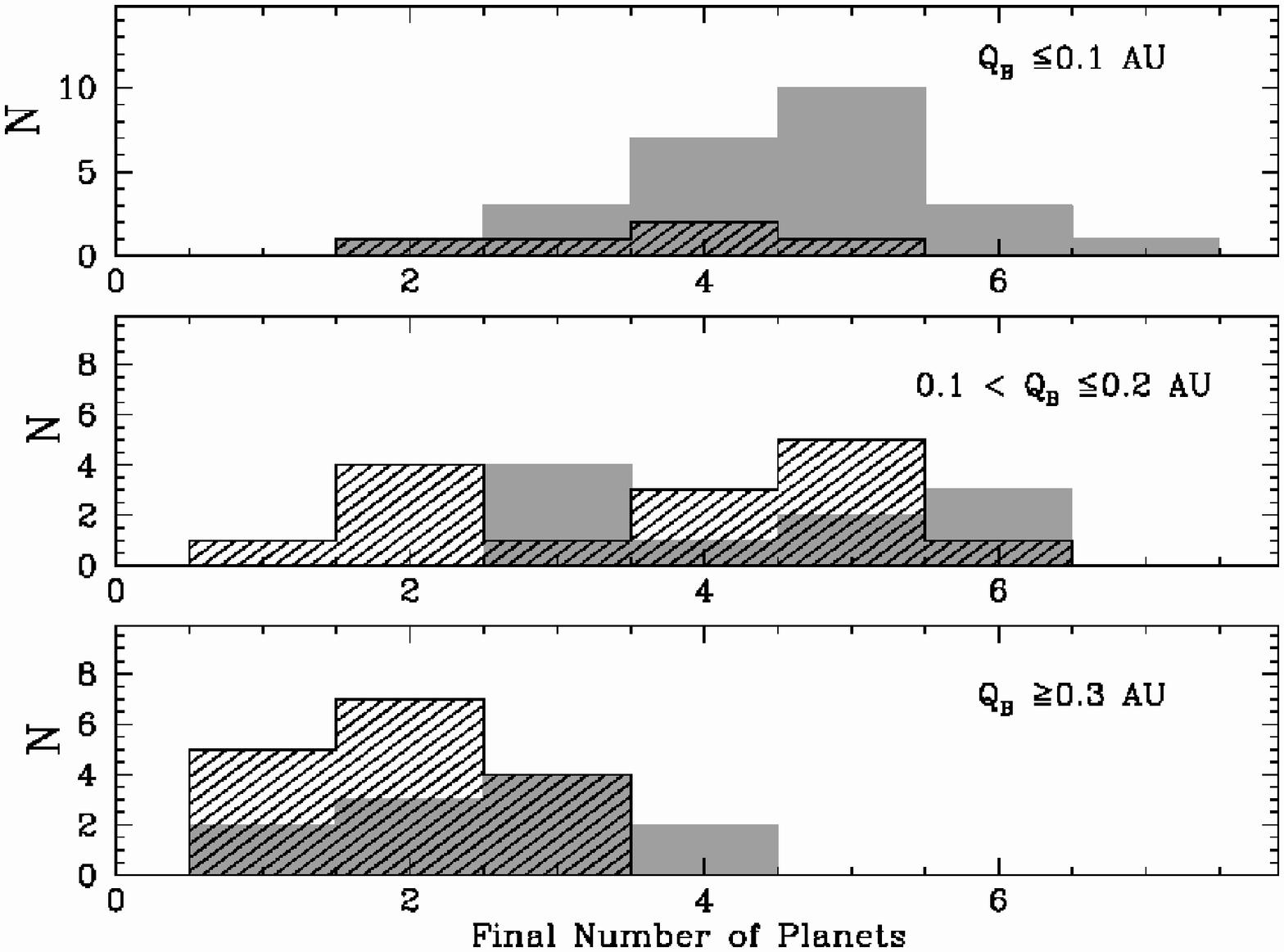}
\caption{Distributions of the semimajor axis of the innermost final
  planet, $a_i$, formed in binary star systems with $Q_B \leq$ 0.1 AU
  (top panel), 0.1 $< Q_B \leq$ 0.2 AU (middle panel), and $Q_B \geq$
  0.3 AU (bottom panel).  The bar types correspond
  to the different sets of runs as described in Figure 10.}

\end{figure}

\begin{figure}[!t]\centering 
\includegraphics[angle=270, width=1\columnwidth]{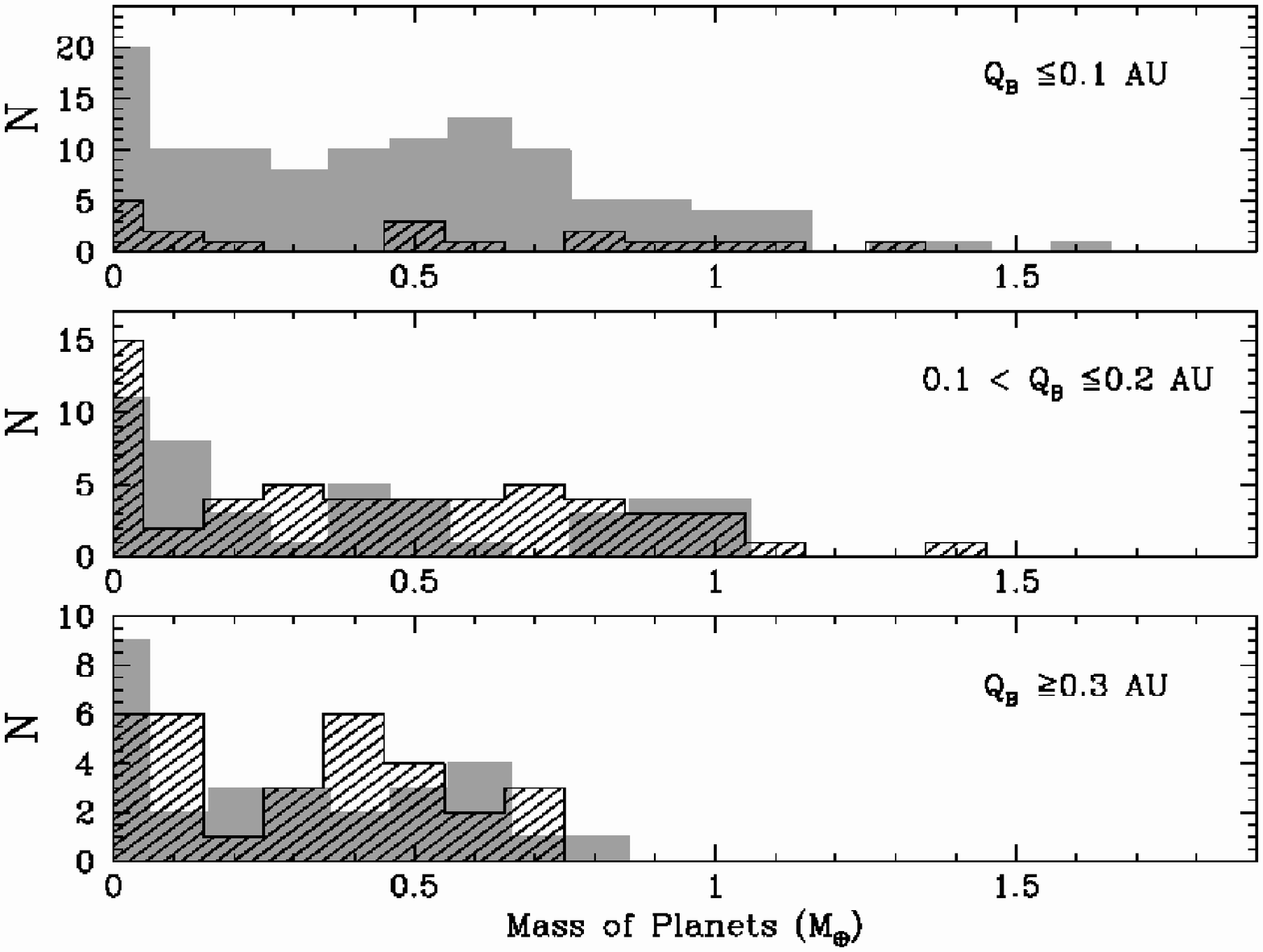}
\caption{Distributions of the semimajor axis of the innermost final
  planet, $a_i$, formed in binary star systems with $Q_B \leq$ 0.1 AU
  (top panel), 0.1 $< Q_B \leq$ 0.2 AU (middle panel), and $Q_B \geq$
  0.3 AU (bottom panel).  The bar types correspond
  to the different sets of runs as described in Figure 10.}
\end{figure}

\printindex

\begin{thebibliography}{99.}

\bibitem{duq91} Duquennoy, A., \& Mayor, M. 1991, A\&A, 248, 485


\bibitem{mat00} Mathieu, R. D., Ghez, A. M., Jensen, E. L. N., \&
  Simon M. 2000, in Protostars and Planets IV, ed. V. Mannings,
  A. P. Boss, \& S. S. Russell (Tucson: Univ. of Arizona Press), 703

\bibitem{bod00} Bodenheimer, P., Hubickyj, O., \& Lissauer, J. J. 2000, Icarus, 143, 2


\bibitem{saf69} Safronov, V. S., 1969. Evolution of
  the Protoplanetary Cloud and Formation of the Earth and the Planets
  (Moscow: Nauka Press).

\bibitem{lis93} Lissauer, J. J. 1993, Ann. Rev. Astron. Astrophys., 31, 129

\bibitem{cha02} Chambers, J. E., Quintana, E. V., Duncan, M. J., \& Lissauer, J. J.  2002, AJ, 123, 2884

\bibitem{qui02} Quintana, E. V., Lissauer, J. J., Chambers, J. E., \& Duncan, M. J. 2002, ApJ, 576, 982


\bibitem{qui03} Quintana, E. V. 2003, in ASP
  Conf. Ser. 294, Scientific Frontiers in Research on Extrasolar
  Planets, ed.  Deming, D., \& Seager, S. (San Francisco: ASP), 319

\bibitem{qui04} Quintana, E. V. 2004,  Planet
  Formation in Binary Star Systems, Thesis (Ph.D.), University of
  Michigan, Ann Arbor (Ann Arbor: UMI Company)

\bibitem{lis04} Lissauer, J. J., Quintana,
  E. V., Chambers, J. E., Duncan, M. J., \& Adams, F. C.  2004,
  RevMexAA, 22, 99


\bibitem{qui06} Quintana, E. V., \& Lissauer, J. J. 2006, Icarus, 185, 1

\bibitem{qui07} Quintana, E. V., F. C. Adams, J. J. Lissauer, \& J. E. Chambers 2007, Ap.J., 660, 807


\bibitem{cha01} Chambers, J. E. 2001, Icarus, 152, 205


\bibitem{end01} Endl, M., Kurster, M., Els, S.,
Hatzes, A. P., \& Cochran, W. D. 2001, A\&A, 374, 675

\bibitem{wie97} Wiegert, P. A., \& Holman, M. J. 1997, AJ, 113, 1445


\bibitem{dav03} David, E., Quintana, E. V.,
  Fatuzzo, M., \& Adams, F. C. 2003, PASP, 115, 825


\bibitem{fat06} Fatuzzo, M., Adams, F. C., Gauvin, R., \& Proszkow, E. M. 2006, PASP, 118, 1510



\end{thebibliography}
\end{document}